\documentclass[12pt,preprint,flushrt]{aastex}

\begin{document}
\newcommand{\etal}{{\it et al.}}

\include{idl}

\title{{\it CHANDRA} Observations of V407 Vul: Confirmation of the
Spin-up} \author{Tod E. Strohmayer} \affil{Laboratory for High Energy
Astrophysics, NASA's Goddard Space Flight Center, Greenbelt, MD 20771;
stroh@clarence.gsfc.nasa.gov}

\begin{abstract}

V407 Vul (RX J1914.4+2456) is a candidate double-degenerate binary
with a putative 1.756 mHz (9.5 min) orbital frequency. In a previous
timing study using archival ROSAT and ASCA data we reported evidence
for an increase of this frequency at a rate consistent with
expectations for gravitational radiation from a detached ultracompact
binary system.  Here we report the results of new {\it Chandra} timing
observations which confirm the previous indications of spin-up of the
X-ray frequency, and provide much tighter constraints on the frequency
derivative, $\dot\nu$. We obtained with {\it Chandra} a total of 90
ksec of exposure in two epochs separated in time by 11.5 months. The
total time span of the archival ROSAT, ASCA and new {\it Chandra} data
is now $\approx 10.5$ years. This more than doubles the interval
spanned by the ROSAT and ASCA data alone, providing much greater
sensitivity to a frequency derivative. With the addition of the
Chandra data an increasing frequency is unavoidable, and the mean
$\dot\nu$ is $7.0 \pm 0.8 \times 10^{-18}$ Hz s$^{-1}$. Although a
long-term spin-up trend is confirmed, there is excess variance in the
phase timing residuals, perhaps indicative of shorter timescale torque
fluctuations or phase instability associated with the source of the
X-ray flux. Power spectral searches for periods longward of the 9.5
minute period do not find any significant modulations, however, the
sensitivity of searches in this frequency range are somewhat
compromised by the dithering of the Chandra attitude. The observed
spin-up is of a magnitude consistent with that expected from
gravitational radiation decay, however, the factor of $\approx 3$
variations in flux combined with the timing noise could conceivably
result from accretion-induced spin-up of a white dwarf.  Continued
monitoring to explore correlations of torque with X-ray flux could
provide a further test of this hypothesis.

\end{abstract}


\section{Introduction}

Ultracompact binary systems could provide a promising means to
observe directly the influence of gravitational radiation on orbital
evolution. Moreover, such systems would be ideal sources for detection
with spaced based gravitational radiation observatories (such as the
planned NASA/ESA LISA mission), opening up the possibility for
detailed studies of compact interacting binaries.

In recent years a pair of candidate ultracompact systems; V407 Vul
(also known as RX J1914.4+2456) and RX J0806+1527 (hereafter J0806)
have been proposed.  These objects were first discovered by ROSAT
(Motch et al.  1996; Israel et al. 1999; Beuermann et al. 1999), and
initially were suggested to be members of a ``soft'' class of
Intermediate Polars (IPs), with the X-ray periods of 569 and 321 s,
respectively, representing the putative spin periods of the accreting
white dwarfs.  Since their discovery extensive follow-up observations
have identified the optical counterparts (Ramsay et al. 2000; Israel
et al. 2002; Ramsay, Hakala \& Cropper 2002). Their soft X-ray
spectra, the shape of the X-ray modulation, the phasing of the X-ray
and optical modulations, the lack of additional longer periods, and
the lack of strong optical emission lines have all called into
question their IP credentials (for a discussion see Cropper et
al. 2003). However, Norton, Haswell \& Wynn (2004) have argued that an
IP interpretation is still plausible if the systems are stream-fed,
pole-switching accretors (ie. no accretion disk), and are viewed from
a nearly face-on geometry. 

It was Cropper et al. (1998) who first suggested that V407 Vul might
be a double-degenerate compact binary.  They proposed a synchronized,
magnetic accretor (polar-like) model with accretion powering the X-ray
flux. A non-magnetic variant was subsequently proposed by Marsh \&
Steeghs (2002).  In this Algol-like model, the accretion stream
impacts directly onto a non-magnetic primary, and the spins are not
necessarily synchronized with the orbit.  An interesting alternative
not requiring accretion was proposed by Wu et al. (2002). They
suggested a unipolar inductor model, analogous to the Jupiter - Io
system (Clarke et al.  1996). If these systems are indeed compact, and
thus the observed X-ray period is the orbital period, then an
important discriminating factor is the magnitude and sign of the
orbital evolution. If the systems are accreting stably from degenerate
donors, the expected evolution is for the orbit frequency to
decrease. In a previous study we (Strohmayer 2002) used archival ROSAT
and ASCA data to explore the evolution of the 1.756 mHz X-ray
frequency of V407 Vul over an $\approx 5$ yr time period, and found
evidence for a positive frequency derivative, $\dot\nu$, with a
magnitude consistent with simple expectations for gravitational
radiation induced decay of a circular orbit. Since a measurement of
the frequency evolution places severe constraints on possible models,
it is crucial to confirm the initial indications of orbital decay and
place tighter constraints on $\dot\nu$.  In this paper we present the
results of new {\it Chandra} observations which confirm an increase in
the X-ray frequency and allow us to place much tighter limits on
$\dot\nu$. In \S 2 we describe the {\it Chandra} observations and the
data extraction and analysis.  In \S 3 we discuss our phase coherent
timing study, and we show that the inclusion of the {\it Chandra} data
conclusively indicates a positive $\dot\nu = 7.0 \pm 0.8 \times
10^{-18}$ Hz s$^{-1}$. We also discuss the flux variability of the
source and excess variance (timing noise) in the phase residuals. In
\S 4 we discuss the implications of our findings for the nature of
V407 Vul. We conclude in \S 5 with a brief summary and goals for
future observations.

\section{Data Extraction}

We obtained a total of $\approx 90$ ksec of exposure on V407 Vul with
{\it Chandra} in February, November, December 2003, and January, 2004.
A summary of these observations is given in Table 1.  We observed with
ACIS-S in Continuous Clocking (CC) mode in order to mitigate the
effects of pile-up. To maximize the soft photon response the aimpoint
was on the backside illuminated (S3) chip. Preliminary spectral
analysis shows that virtually all the source photons have energies
less than 1 keV. For the purposes of our work reported here we used
only $< 1$ keV photons.  Details of the {\it Chandra} spectroscopy
will be presented in a future paper.

To prepare the data for a precise timing analysis we first corrected
the detector read out times to arrival times following the CXC
analysis thread on timing with CC mode data. We then corrected the
arrival times to the solar system barycenter using the CIAO tool {\it
axbary}. We used the same source coordinates employed for analysis of
the ROSAT and ASCA data (see Strohmayer 2002). This produces a set of
photon arrival times in the barycentric dynamical time system
(TDB). The ACIS/CC mode produces a one-dimensional ``image'' of the
portion of the sky exposed to the detectors. A realization of this
image for the February, 2003 observations is shown in Figure 1. V407
Vul is the strong peak in the ``image.'' We extracted only events from
within the source peak for our timing study.  Figure 2 shows a portion
of the lightcurve produced from the source extracted events, and
demonstrates that {\it Chandra} easily detects individual pulses from
the source. We carried out this procedure on all the data and obtained
a total of 29,383 {\it Chandra} events for our timing analysis.

\section{Results}

\subsection{Coherent Timing Analysis}

We performed our coherent timing studies using the $Z_n^2$
statistic. Since the method is described in our earlier paper we do
not repeat all the details here (see Strohmayer 2002).  As a first
step we calculated two $Z_3^2$ power spectra. The first uses just the
earlier ROSAT and ASCA data, while the second uses only the new {\it
Chandra} data. This procedure allows us to compare the frequencies
measured at two widely spaced epochs. A comparison of the two then
determines the nature of any long term trend in the frequency
evolution. This is perhaps the simplest possible analysis that one can
do to search for a frequency change. The results are shown in Figures
3 and 4.  Figure 3 shows the $Z_3^2$ power spectra for the two
different epochs, ROSAT + ASCA (top), and {\it Chandra} (bottom). In
each case we plot the difference, $\Delta Z_3^2$, between the peak
value of $Z_3^2$ and the values at other frequencies. In the absence
of a signal, the $Z_3^2$ statistic has noise properties given by the
$\chi^2$ distribution with $2\times 3 = 6$ degrees of freedom
(dof). The ``forest'' of minima in each plot results from the
relatively sparse data sampling. Note, however, that for each dataset
the best frequency {\it can} be identified unambiguously (the one with
$\Delta Z_3^2 = 0$). This is crucially important, as correctly pointed
out by Woudt \& Warner (2004), who cautioned that frequency evolution
claims for RX J0806.3+1527 and V407 Vul (see Hakala et al. 2003;
Strohmayer 2003; Strohmayer 2002) might be compromised by an inability
to identify the correct frequency at a given epoch. For the ROSAT +
ASCA data the next-best alias peak is at $\Delta Z_3^2 \approx 1300$
from the best frequency. This is an enormously significant difference
in $\Delta Z_3^2$. The probability that this alias could be the true
frequency, compared to a frequency with a $Z_3^2$ value larger by
1300, is vanishingly small. The same is true for the {\it Chandra}
data, for which the nearest alias peak has a $\Delta Z_3^2 \approx
696$. Although a smaller difference than for the ROSAT + ASCA data,
the probability of this being the true frequency is still incredibly
small. As an example, the probability to obtain a difference in
$\Delta Z_3^2$ of 100.0 (much smaller than the observed values) purely
by chance is $ < 3 \times 10^{-19}$.

Having shown that at each epoch we can unambiguously identify the
correct frequency, we can now compare these two frequencies and test
whether the frequency has changed. Figure 4 shows a zoom-in around the
best ROSAT + ASCA (dashed) and {\it Chandra} (solid) frequencies, and
clearly demonstrates that the frequency is higher in the {\it Chandra}
epoch. The increase in frequency is about $2$ nHz. A simple estimate
of the mean frequency derivative can be obtained by simply dividing
the measured frequency increase by the time interval between the mean
epoch of the two datasets.  This gives a value $<\dot\nu> = 2 \times
10^{-9} \ {\rm Hz} / 7.6 \ {\rm yr} = 8.34 \times 10^{-18}$ Hz
s$^{-1}$, which is consistent with the limits on $\dot\nu$ deduced
from our earlier study as well as the more detailed analysis described
below.

We next performed a coherent, phase connected timing analysis to
jointly constrain $\nu_0$ and $\dot\nu$. As in our previous study we
carried out a grid search by calculating $\chi^2$ at each $\nu_0$,
$\dot\nu$ pair (see Strohmayer 2002 for details).  The results are
summarized in Figures 5, 6 and 7.  Figure 5 shows two pairs of
$\Delta\chi^2$ confidence contours.  The large dashed contours
represent the constraints derived only from the ROSAT + ASCA data, and
the solid contours represent the best solution obtained with the
inclusion of the new {\it Chandra} data. The analysis definitively
favors a positive $\dot\nu = 7 \pm 0.8 \times 10^{-18}$ Hz s$^{-1}$.
The parameters of our best timing ephemeris are given in Table 1.

Although the presence of a long term spin up trend in the data is now
indisputable, the minimum $\chi^2$ value of 390 (with 98 degrees of
freedom) for our best fit is formally high.  Because of this we first
scaled the $\chi^2$ values for the whole dataset (including {\it
Chandra}) by the ratio $390/98 \approx 4$ before plotting the
confidence contours. This gives a conservative estimate of the
confidence region for $\nu$ and $\dot\nu$. Figure 6 shows two
representations of the phase measurements. The phase residuals
obtained with $\dot\nu = 0$ and our best constant frequency, $\nu_0$,
for all the data are shown in the top panel.  In this representation a
positive $\dot\nu$ (spin-up) will show up as a quadratic trend with
the parabola opening downward. One can clearly pick out such a trend
with the eye. The phase residuals computed with our best ephemeris
(ie. including the best non-zero value of $\dot\nu$) are shown in the
bottom panel. Figure 7 is essentially identical to Figure 6 except the
time gaps have been removed so that all the individual residuals can
be more easily seen. Note the much higher statistical quality of the
Chandra data.

The excess $\chi^2$ value is indicative of some additional ``timing
noise'' in the system. Although the rms deviations are qualitatively
similar for all the data, the much higher statistical quality of the
{\it Chandra} data contributes substantially to the excess. The
presence of timing noise could in principle result from fluctuations
in the spin up torque, or perhaps from modest changes in the profile
of the pulsed emission. By monitoring the character of such
fluctuations it should be possible to further constrain models for the
spin up torque. We discuss this further below. Since fluctuations of
both sign appear, and over a range of timescales, it seems extremely
unlikely that the overall spin up trend could somehow result from a
random superposition of fluctuations with just the right amplitude and
phasing.Using our best timing ephemeris we have calculated folded
pulse profiles for the ROSAT + ASCA and new {\it Chandra} data. The
results are shown in Figure 8. The {\it Chandra} profile has been
displaced vertically by 600 counts for clarity.  Phase zero in the
plot corresponds to MJD 49257.533373137 (TDB). The pulse profile
appears to be rather stable over the past decade.

\subsection{Flux Variability}

An important constraint on models for V407 Vul is the present lack of
any detectable periodicities with timescales longer than the 1.756 mHz
X-ray modulation.  If the system were in fact an IP, then one might
expect that the longer, unseen orbital period could eventually be
detected with sensitive observations. The {\it Chandra} data are
sensitive enough to explore variability on timescales longer than the
X-ray pulsation.  We did this by computing a lightcurve using our best
fit period as the time bin size.  The resulting lightcurve is shown in
Figure 9. In this plot we have suppressed the time gaps. Individual
observations are separated by the vertical dotted lines. Epochs for
each of the five observations can be found in Table 1. The data show
evidence for both long term variations in the flux as well as some
pulse to pulse variations.  For example, the flux during our
next-to-last pointing was a factor of 3 higher than in the initial
pointing.  The longest pointing (2nd panel from the left) shows
evidence for $30 - 50 \%$ variations in the flux within a few pulse
periods.

Our 35.4 ksec pointing at V407 Vul is the longest contiguous
observation of the source to date. This, combined with the higher
signal to noise ratio in the {\it Chandra} data compared with ROSAT or
ASCA, allows, in principle, more sensitive searches for a putative IP
orbital period.  To search for periods longward of the 1.756 mHz
frequency we computed Fourier power spectra of each of our
observations.  We do not find any robust periodicities which are
present in all of our observations.  In the power spectrum of the
longest pointing a weak sideband structure accompanies the fundamental
and first two harmonics of the 1.756 mHz frequency. The low frequency
region of this power spectrum is shown in Figure 10. The strongest of
these peaks are the upper sidebands to the fundamental and first
harmonic peaks at 1.756 mHz and 3.512 mHz. These sidebands are
separated in frequency from their parent peaks by 0.4825 mHz, and have
Leahy normalized powers of 74.8 and 52.5. There also appear to be
weaker lower sidebands accompanying the first and second harmonic
peaks at 3.512 mHz and 5.28 mHz respectively. These lower sidebands
are separated in frequency by 0.273 mHz from their respective
harmonic peaks.  The upper sideband frequency separation is close to,
but significantly different from, twice the lower sideband separation.

Although such a sideband structure could conceivably be intrinsic to
the source, and thus could suggest a longer, previously unseen period
in the system, we think a more plausible explanation is that these are
spurious signals most likely introduced by the dithering of the {\it
Chandra} pointing position. {\it Chandra} dithers in a Lissajous
pattern with different periods in the horizontal and vertical
directions. For ACIS, the dither periods are 1000 and 707 seconds in
the Y and Z directions, respectively. Indeed, there is a modest peak
at a 1000 s period in the power spectrum of our longest
observation. It seems likely that beating between the strong 1.756 mHz
modulation and the dither periods could introduce sideband
modulations. Moreover, since the strength of power spectral features
is proportional to the counting rate, we would expect an astrophysical
signal, if robust, to produce a greater signal in our brightest
observation, yet there is no evidence for the sideband structure in
the observation at the highest counting rate.  Based on these
arguments we do not think there is yet sufficient evidence to claim
detection of a longer period.  Because of the systematic issues
associated with dithering {\it Chandra}, it seems likely that XMM
observations will provide more definitive results with regard to
searches for an unseen orbital period in V407 Vul.

\section{Implications for the Nature of V407 Vul}

Our {\it Chandra} data definitively shows that the 1.756 mHz X-ray
modulation of V407 Vul is increasing at a mean rate of $\approx 7
\times 10^{-18}$ Hz s$^{-1}$. This rate is consistent with what would
be expected if the X-ray modulation is the orbital period of an
ultracompact system and it's orbit were decaying via emission of
gravitational radiation (see Strohmayer 2002).  Based on the observed
spin-up there would appear to be two remaining alternatives regarding
the nature of the source. 1) An ultracompact but detached system
powered by electrical energy, as described by Wu et al. (2002), or 2)
a low-inclination IP system in which the observed X-ray modulation is
effectively the white dwarf spin, and the X-ray flux and observed
spin-up are accretion driven (Norton, Haswell \& Wynn 2004). Other
ultracompact scenarios appear to be ruled out by the observed spin-up
and/or the short orbital period (see Strohmayer 2002; Cropper et
al. 2003).

In the IP scenario the white dwarf accretion is stream-fed and
switches from pole to pole and back again on the synodic period. An
accretion disk would form if the magnetospheric radius is less than
the circularization radius for the accretion stream. In order not to
form a disk and for V407 Vul to be stream-fed the orbital period must
satisfy $P_{orb} < 2.37$ hr (Norton et al. 2004).  If this scenario is
correct, then the secular spin up is due to accretion onto the white
dwarf, and other aspects of the observations, for example, the long
term flux variability and the phase timing noise, may be more easily
understood in terms of variations in the mass accretion rate. This
identification would also preclude the need to invoke models perceived
by some as ``exotic'' (such as the Wu et al. (2002) electric star
scenario).

However, some intriguing questions remain, for example, the soft X-ray
spectrum, the absence (or at least weakness) of optical emission
lines, and the infrared photometry have all been used to argue against
an IP identification (see Cropper et al. 2003 for a discussion).
Norton et al. (2004) have argued that the relatively high mass
accretion rates, of order $10^{-9} M_{\odot}$ yr$^{-1}$, required for
accretion to power the observed X-ray flux, as well as the nature of
stream-fed accretion may mitigate most of these perceived
difficulties.  For example, a highly optically thick stream may mask
the emission lines, and if the accretion is ``buried'' in the white
dwarf surface layers, then it may manifest itself with predominantly
soft blackbody emission. Deeper optical spectroscopy would likely
provide important new insight.

Accretion-induced spin-up should be testable with continued phase
coherent monitoring.  Comparison of torque fluctuations with X-ray
flux (ie. mass accretion rate) could yield a signature of magnetically
controlled accretion (see for example, Ghosh \& Lamb 1979). Moreover,
if the frequency derivative changes sign that would strongly support
the IP scenario.

\section{Summary}

Our new {\it Chandra} data confirm that the 1.756 mHz X-ray frequency
of RX V407 Vul is increasing at a rate of $\approx 7 \times 10^{-18}$
Hz s$^{-1}$.  Although this rate is consistent with that expected from
gravitational radiation losses in a detached ultracompact binary, an
IP interpretation in the context of accretion onto a spinning white
dwarf cannot yet be strongly ruled out.

In some ways the new {\it Chandra} results have only deepened the
mystery surrounding V407 Vul (and by implication, its sister source,
RX J0806.3+1527). After a decade of observations we still do not know
the nature of the source with any certainty.  However, a number of
future observations could help provide the solutions. Deep pointings
with XMM could probe more sensitively for an unseen orbital period,
and further timing observations with {\it Chandra} will establish
constraints on the dependence of torque fluctuations on X-ray
flux. Finally, if these do not suffice to crack the mystery, a
detection with a spaced-based gravitational radiation observatory
(such as the NASA/ESA LISA mission) would provide definitive evidence
for an ultracompact system.

\acknowledgements

We thank Richard Mushotzky, Zaven Arzoumanian, Jean Swank and Craig
Markwardt for many helpful comments and discussions. We also thank the
referee for his/her detailed comments on the original manuscript, that
helped us to significantly improve the paper. This work made use of
data obtained from the High Energy Astrophysics Science Archive
Research Center (HEASARC) at Goddard Space Flight Center.

\centerline{\bf References}





759 

\noindent{} Beuermann, K., Thomas, H. -C., Reinsch, K., Schwope,
A. D., Trumper, J. \& Voges, W. 1999, A\&A, 347, 47.


\noindent{} Clarke, J. T. et al. 1996, Science, 274, 404

\noindent{} Cropper, M., Ramsay, G., Wu, K. \& Hakala, P. 2003, ASP
Conference Series to be published in Proc. Cape Town Workshop on
magnetic CVs, held Dec 2002, (astro-ph/0302240).

\noindent{} Cropper, M. et al. 1998, MNRAS, 293, L57

\noindent{} Ghosh, P. \& Lamb, F. K. 1979, ApJ, 232, 259.

\noindent{} Hakala, P. et al. 2003, MNRAS, 343, 10L.







\noindent{} Israel, G. L. et al. 1999, A\&A, 349, L1.

\noindent{} Israel, G. L. et al. 2002, A\&A, 386, L13.

\noindent{} Kuulkers, E., Norton, A., Schwope, A. \& Warner, B. 2003, in 
``Compact Stellar X-ray Sources,'' ed. M. van der Klis \& W. H. G. Lewin,
(Cambridge Univ. Press: Cambridge, UK).

\noindent{} Marsh, T. R. \& Steeghs, D. 2002, MNRAS, 331, L7

\noindent{} Motch, C. et al. 1996, A\&A, 307, 459



\noindent{} Norton, A. J., Haswell, C. A. \& Wynn, G. A. 2004, A\&A,
in press, (astro-ph/0206013)


\noindent{} Ramsay, G., Cropper, M., Wu, K., Mason, K.~O., \& Hakala, P.\ 
2000, MNRAS, 311, 75

\noindent{} Ramsay, G., et al. 2002, MNRAS in press, (astro-ph/0202281)

\noindent{} Ramsay, G. Hakala, P. \& Cropper, M. 2002, MNRAS, 332, L7.


\noindent{} Ritter, H. \& Kolb, U. 2003,, A\&A, 404, 301.





\noindent{} Strohmayer, T. E., 2003, ApJ, 593, 39L.

\noindent{} Strohmayer, T. E., 2002, ApJ, 581, 577.





\noindent{} Warner, B. 1995, {\it Cataclysmic Variable Stars}, Cambridge
Univ. Press, Cambridge UK.

\noindent{} Warner, B. 1986, MNRAS, 219, 347.

\noindent{} Woudt, P. A. \& Warner, B. 2003, in Proc. Cape Town Workshop on 
Magnetic CVs, (astro-ph/0310494).

\noindent{} Wu, K., Cropper, M., Ramsay, G. \& Sekiguchi, K. 2002,
MNRAS, 331, 221

\clearpage

\begin{figure}
\begin{center}
 \includegraphics[width=6in, height=6in]{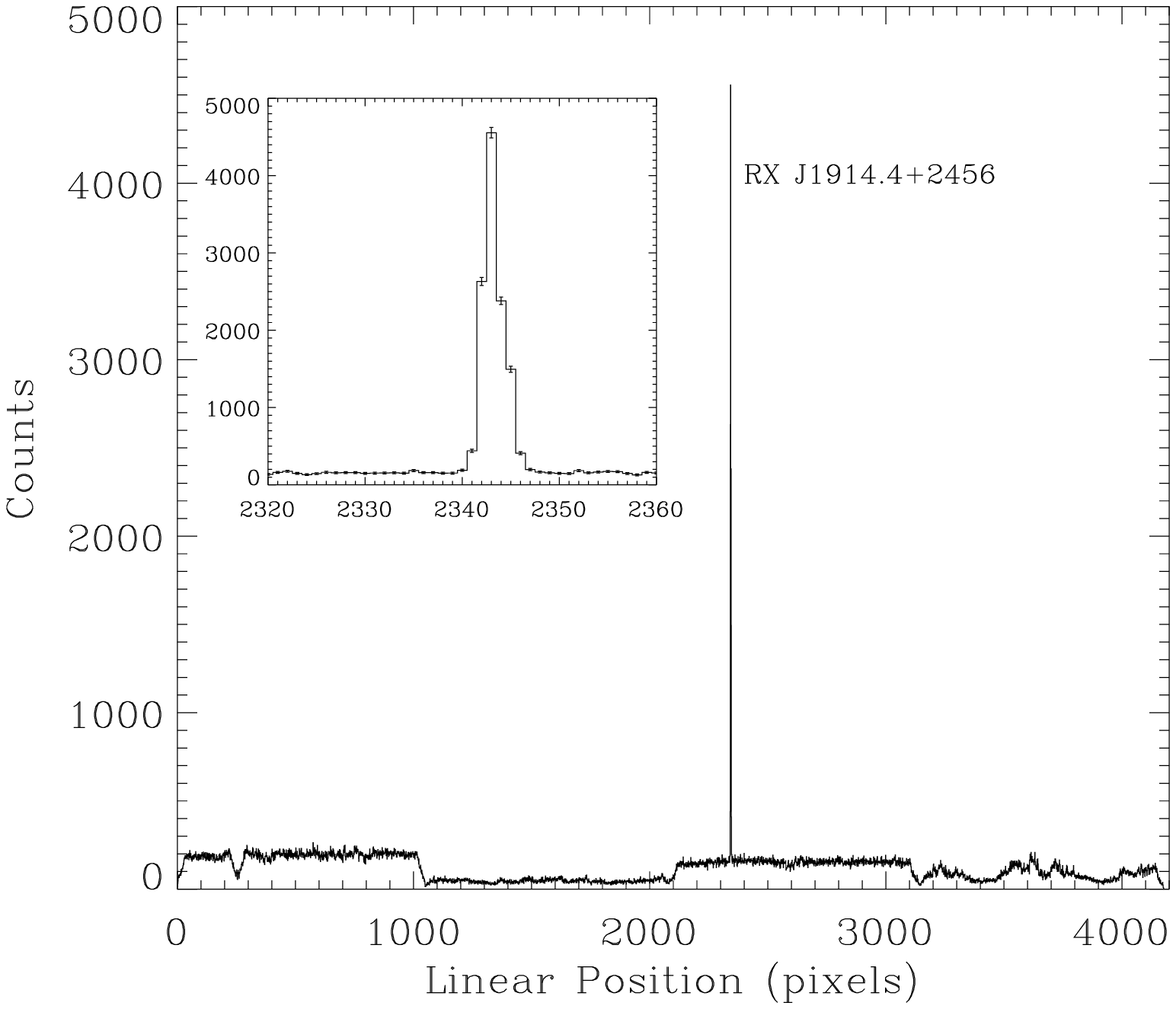}
\end{center}
Figure 1: {\it Chandra} 1-d image of V407 Vul from ACIS-S Continuous
Clocking mode data. The sharp peak contains photons from V407 Vul.
The inset panel shows an expanded view of the peak.
\end{figure}
\clearpage

\begin{figure}
\begin{center}
\includegraphics[width=6in, height=6in]{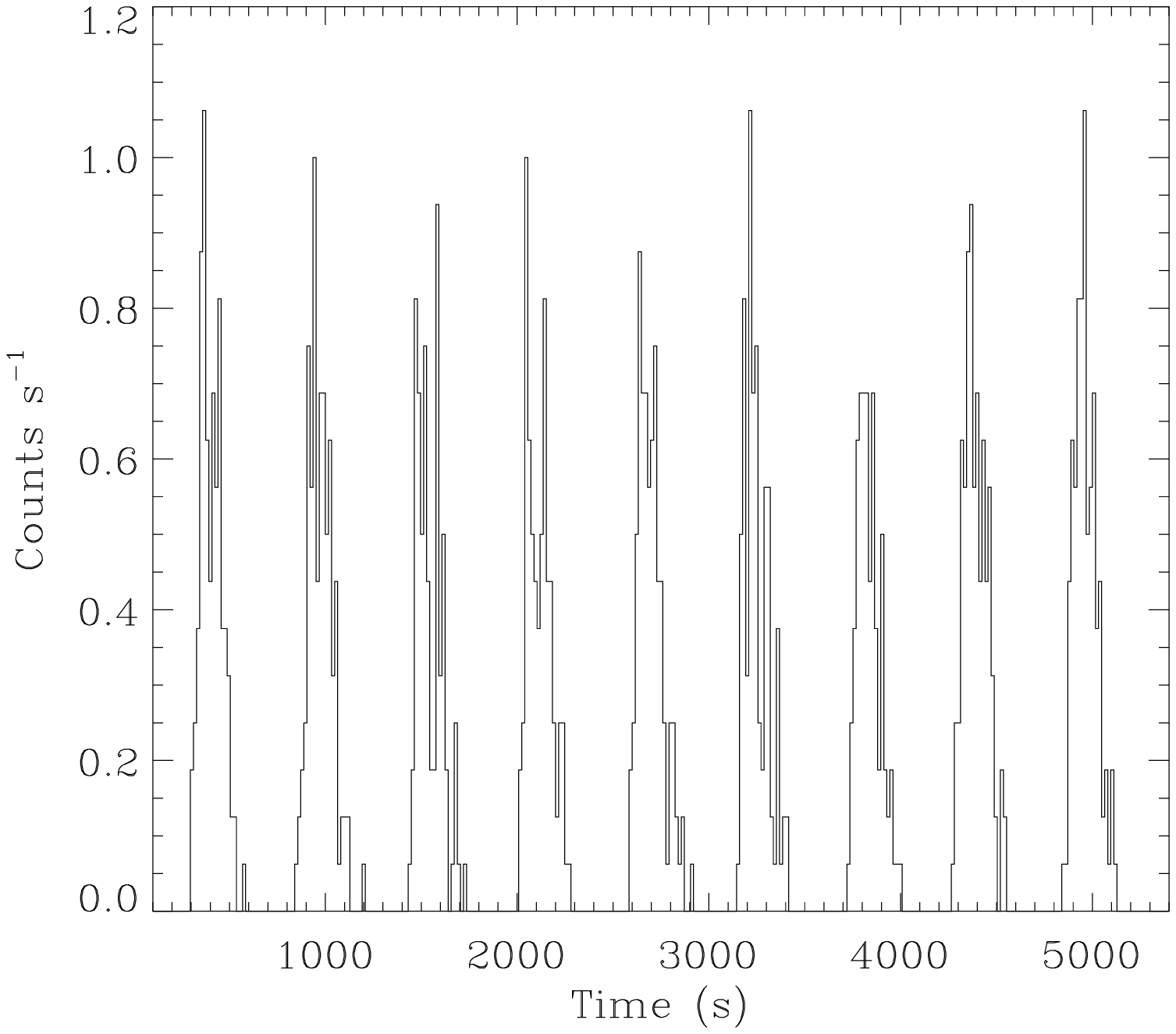}
\end{center}
Figure 2: A lightcurve of a portion of the {\it Chandra} ACIS-S data
from V407 Vul. The binsize is 16 s, and nine individuals pulses are
shown.
\end{figure}
\clearpage

\begin{figure}
\begin{center}
 \includegraphics[width=6in, height=5in]{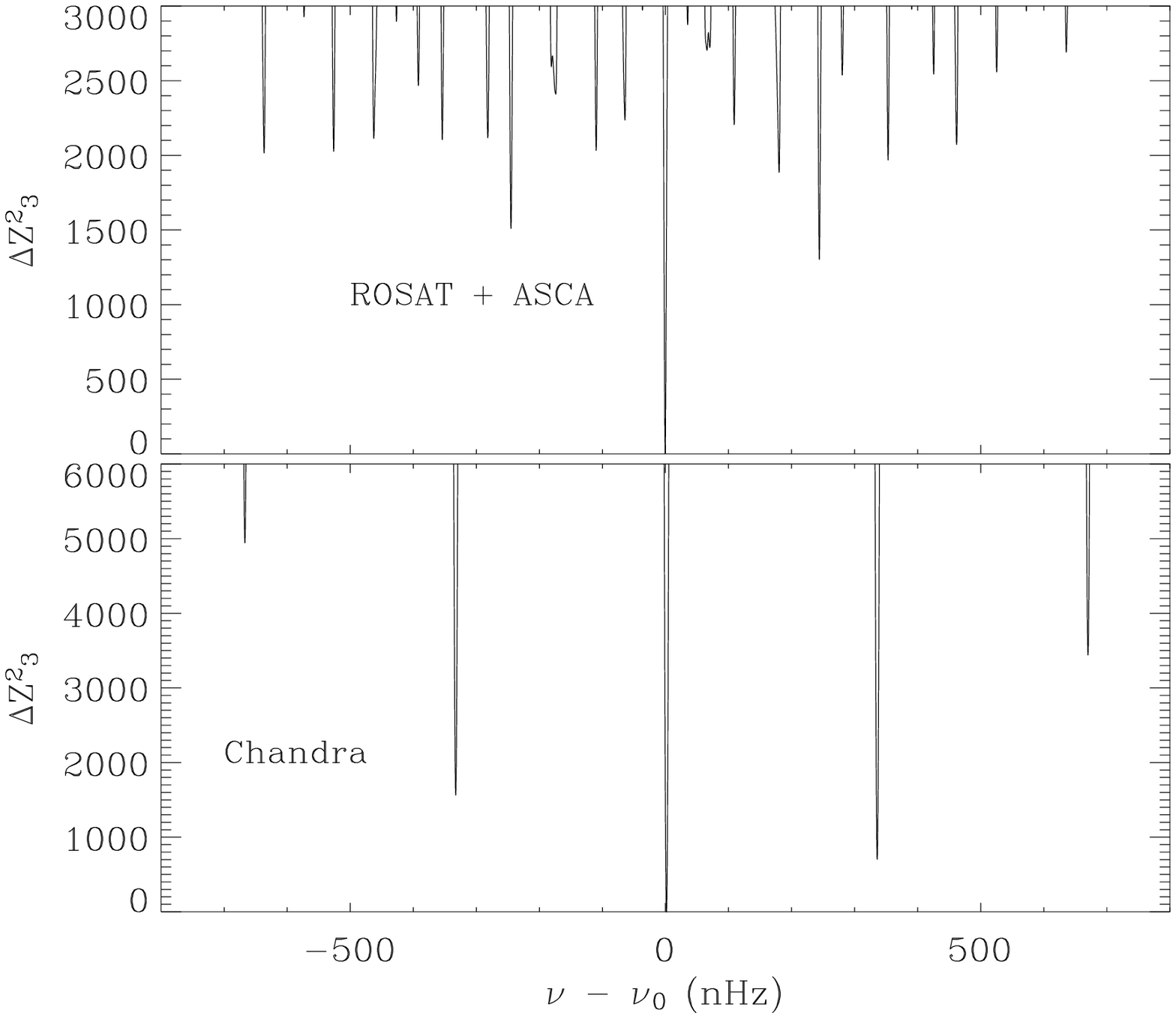}
\end{center}
Figure 3: Best constant frequency measurements for V407 Vul at two
different epochs; the combined ROSAT + ASCA data (top), and the new
{\it Chandra} data (bottom). Shown are the values of $\Delta Z_3^2 =
{\rm max}(Z_3^2) - Z_3^2$ as a function of frequency.  In this
representation the minima denote the best frequency values for each
dataset.  Multiple minima are present due to the inclomplete sampling
of the datasets. In each case, however, the correct frequency peak can
be identified unambiguously (the peaks centered near $\nu - \nu_0 =
0$). All other peaks are at significantly high values of $\Delta
Z_3^2$ to be strongly excluded (see the text for further details). The
reference frequency, $\nu_0$, is $1.7562475 \times 10^{-3}$ Hz.
\end{figure}
\clearpage

\begin{figure}
\begin{center}
 \includegraphics[width=6in,
 height=6in]{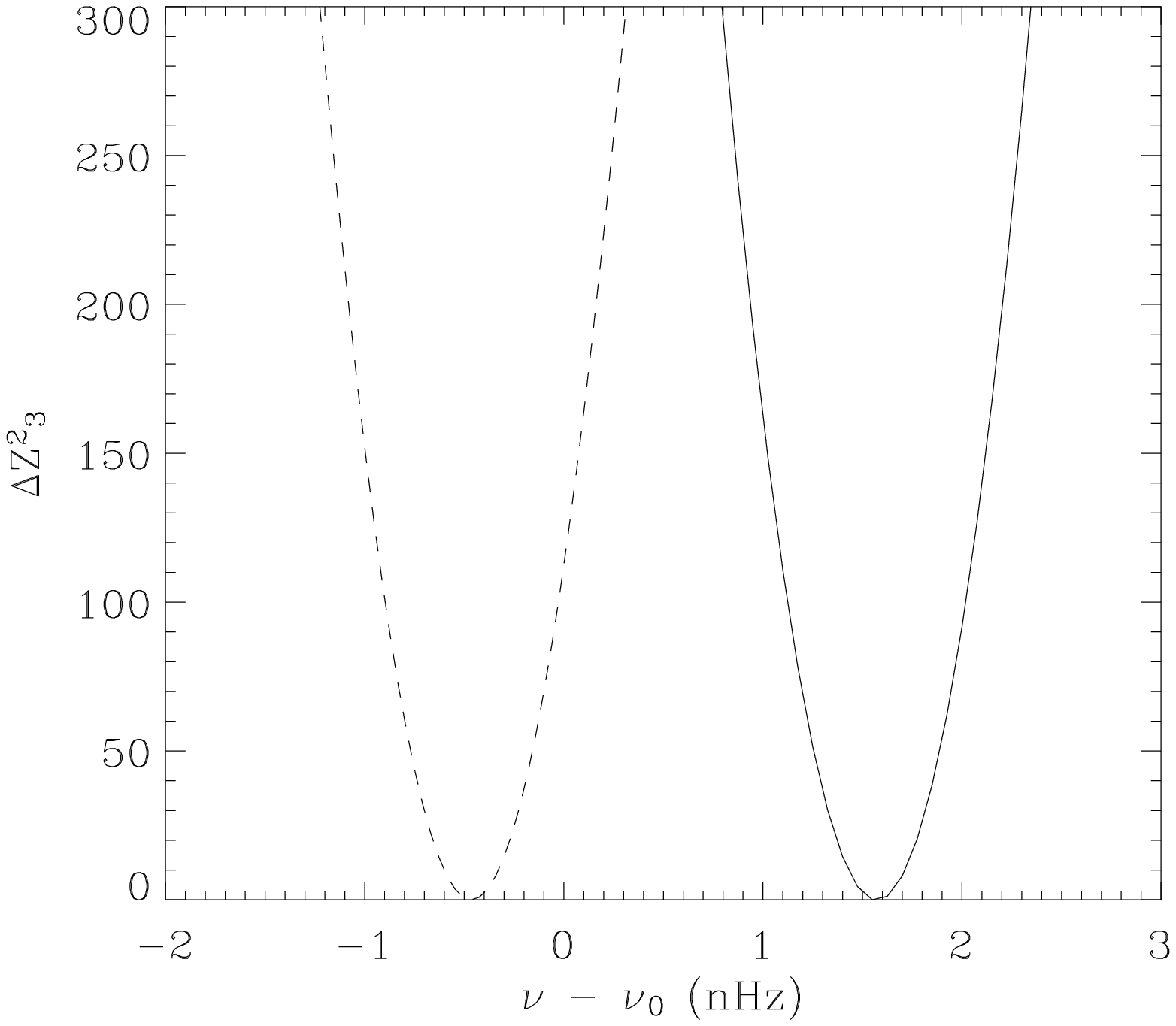}
\end{center}
Figure 4: Same as Figure 3 but now showing an expanded view of the
region around $\nu - \nu_0 = 0$. The best frequencies from the ROSAT +
ASCA (dashed) and Chandra (solid) epochs are shown. The frequency
measured at the more recent {\it Chandra} epoch is clearly higher.
This establishes beyond doubt that the frequency of V407 Vul has
increased over the last 11 years (see the text for further details).
\end{figure}
\clearpage

\begin{figure}
\begin{center}
 \includegraphics[width=6in,height=6in]{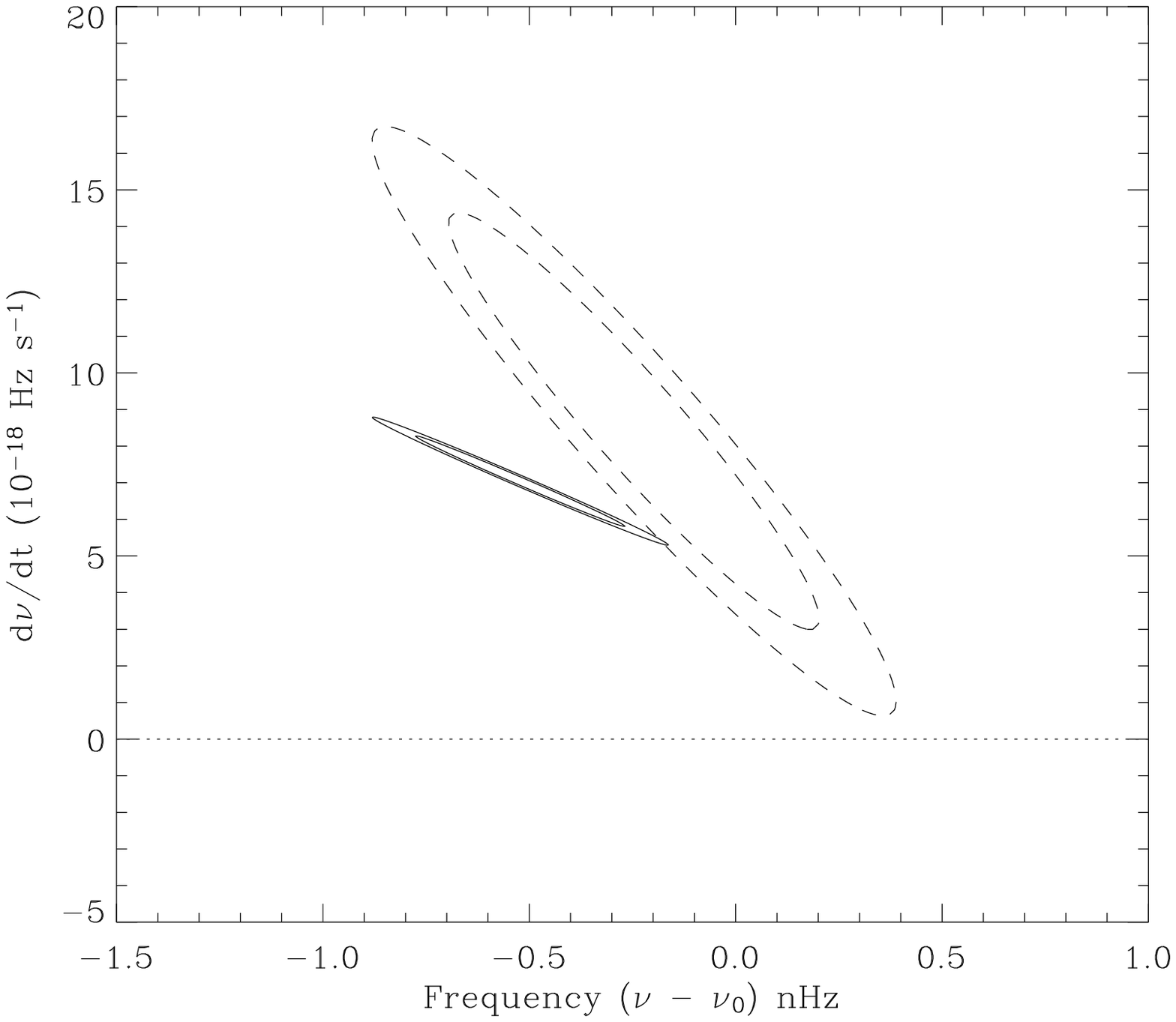}
\end{center}
Figure 5: Constraints on $\nu$ and $\dot\nu$ from our phase timing
analysis of the ROSAT, ASCA and new {\it Chandra} data for V407
Vul. Shown are the 90 and 99\% confidence regions for the ROSAT + ASCA
data (dashed), and the combined dataset (ROSAT, ASCA and {\it
Chandra}).  In plotting the confidence contours for all the data
(solid) we first scaled the $\chi^2$ values by the factor 390/98 (see
text for discussion). The inclusion of the {\it Chandra} data
definitively requires a positive $\dot\nu = 7 \pm 0.8 \times 10^{-18}$
Hz s$^{-1}$.
\end{figure}
\clearpage

\begin{figure}
\begin{center}
 \includegraphics[width=6in, height=6in]{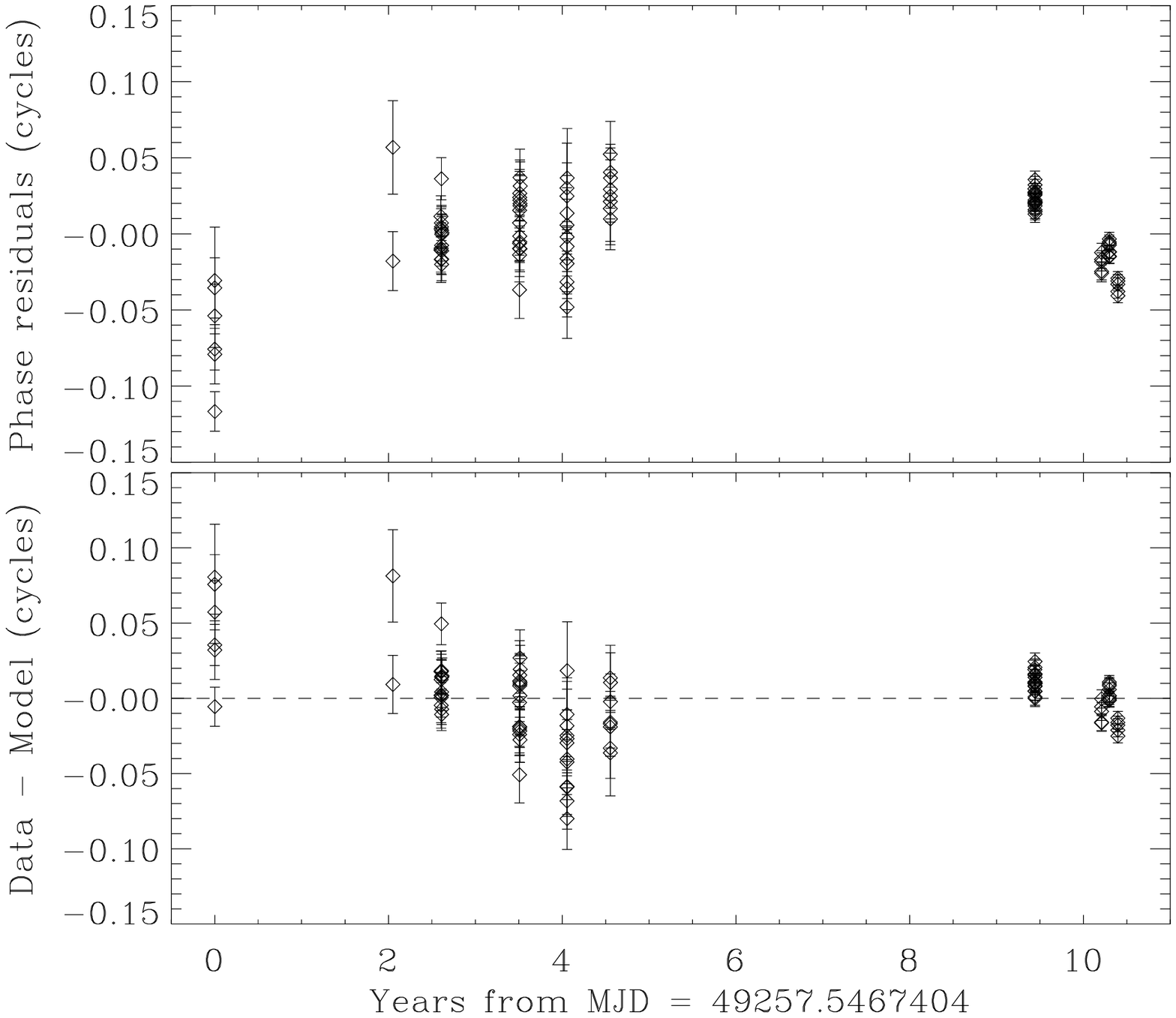}
\end{center}
Figure 6: Phase timing residuals for V407 Vul plotted as a function of
time using a constant frequency phase model (ie. $\dot\nu \equiv 0$
(top), and with the best fitting solution including a positive $\dot
\nu = 7 \times 10^{-18}$ Hz s$^{-1}$ (bottom). In the top panel a
parabolic shape with the parabola opening downward is indicative of
the need for a positive $\dot\nu$. Such a trend can just about be
discerned with the eye. The {\it Chandra} data are those points beyond
year 9.
\end{figure}
\clearpage

\begin{figure}
\begin{center}
 \includegraphics[width=6in, height=6in]{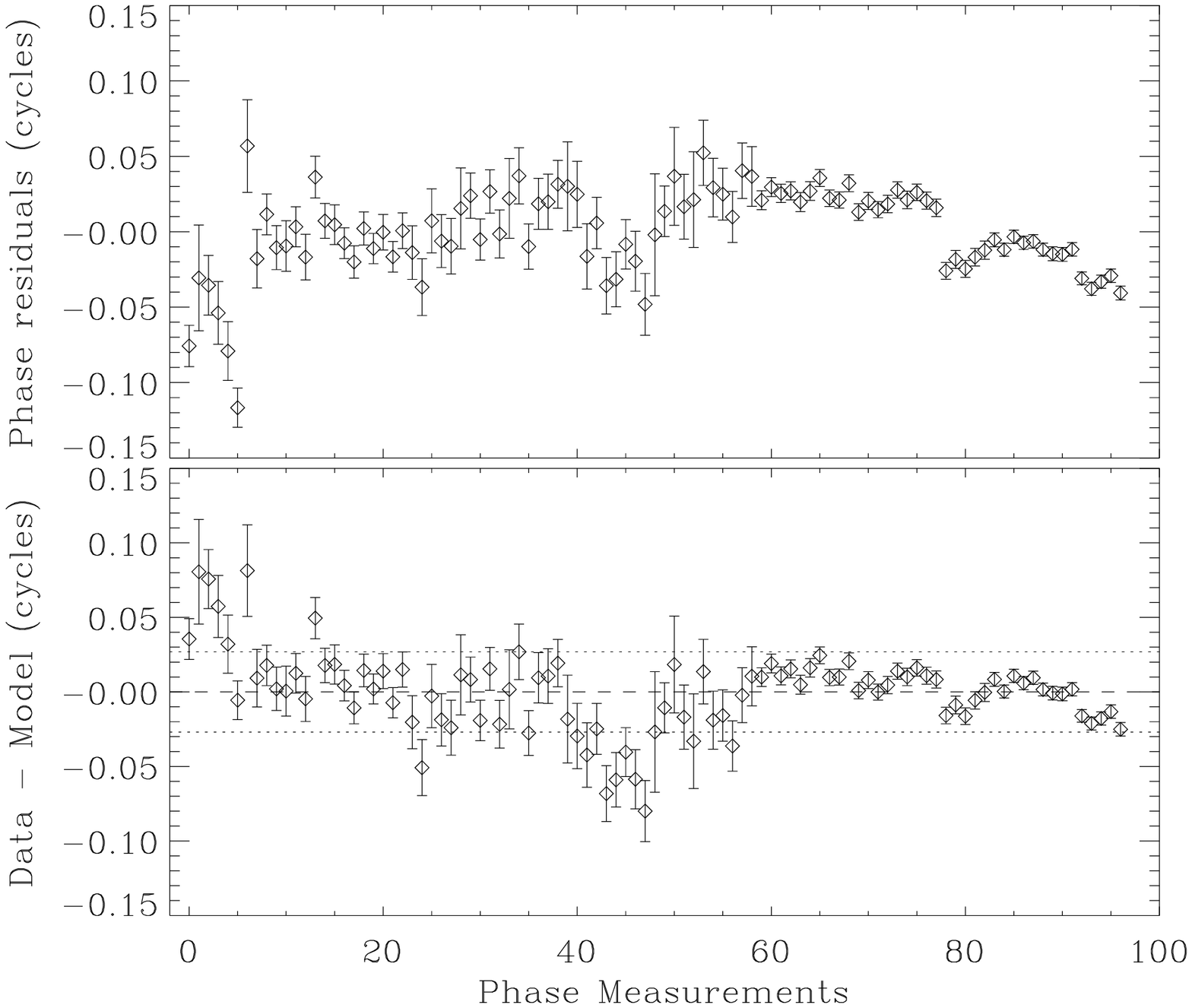}
\end{center}
Figure 7: Same as Figure 6 except the time gaps have been removed so
that the individual phase residuals can be more easily seen. The rms
deviation of the residuals is also shown (dotted lines). The {\it
Chandra} data are those beyond measurment number 58 (note the smaller
error bars).
\end{figure}
\clearpage

\begin{figure}
\begin{center}
 \includegraphics[width=6in, height=6in]{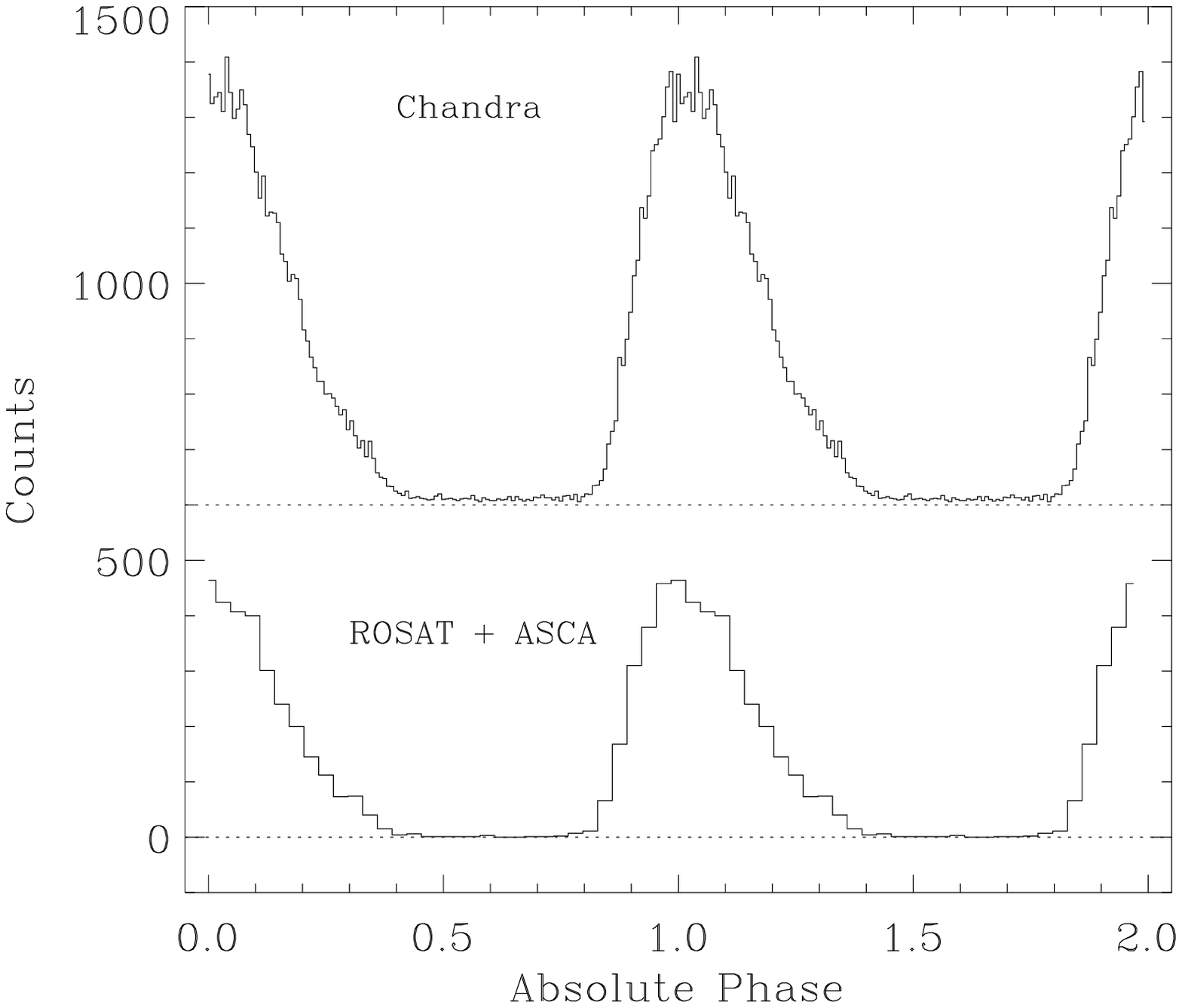}
\end{center}
Figure 8: Folded pulse profiles for the ROSAT + ASCA (bottom) and new
{\it Chandra} data (top) using our best timing ephemeris (see Table
1). The {\it Chandra} profile has been displaced vertically by 600
counts for clarity. Phase zero corresponds to MJD 49257.533373137
(TDB).
\end{figure}
\clearpage

\begin{figure}
\begin{center}
 \includegraphics[width=6in, height=6in]{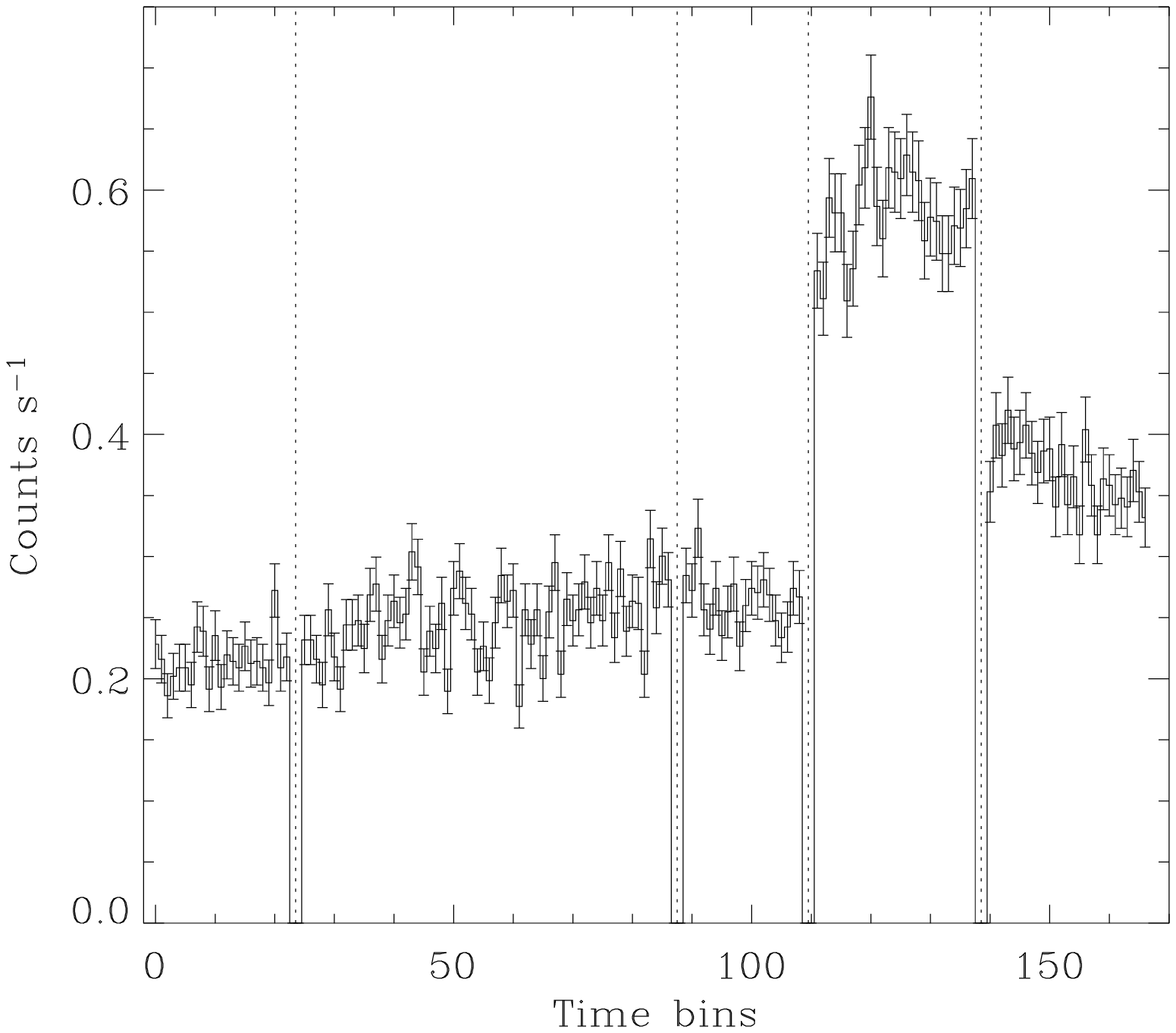}
\end{center}
Figure 9: Flux variability of V407 Vul seen with {\it Chandra} at five
different epochs. The time binsize was set equal to one cycle of the
1.756 mHz pulsation. There is clear evidence for large, long term flux
modulations as well as shorter timescale fluctuations.
\end{figure}
\clearpage

\begin{figure}
\begin{center}
 \includegraphics[width=6in, height=6in]{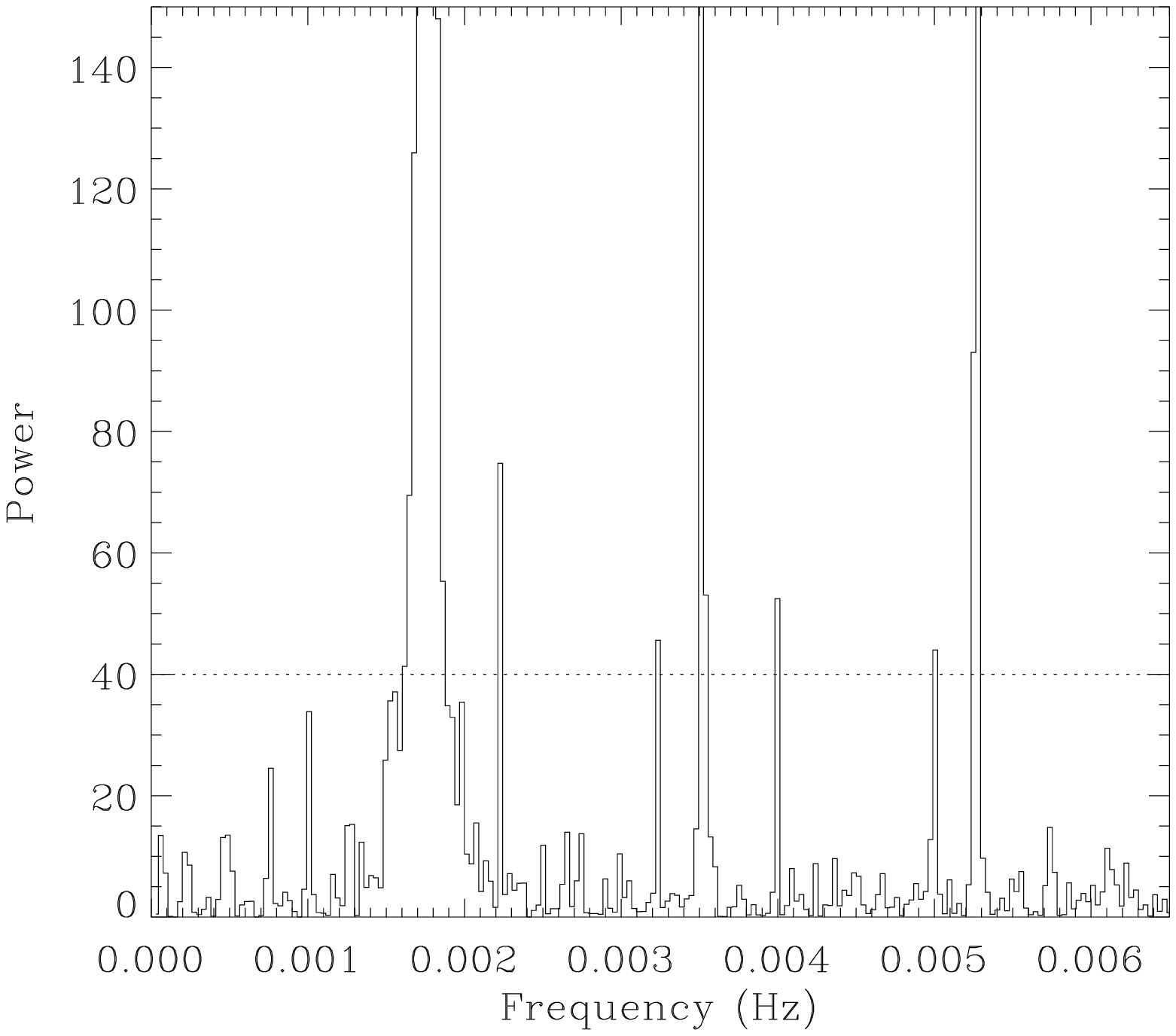}
\end{center}
Figure 10: Low frequency power spectrum of our longest {\it Chandra}
pointing. The power spectrum is normalized such that a pure poisson
noise process would have a mean of 2 and be distributed as $\chi^2$
with two degrees of freedom.  The three peaks that go off-scale are
the fundamental, first and second harmonic of the 1.756 mHz
modulation.  Upper sidebands to the fundamental and first harmonic,
and lower sidebands to the first and second harmonics are present. The
horizontal dotted line indicates the power level with a single trial
chance probability of $\approx 2 \times 10^{-9}$. We suspect the
sidebands are an artifact introduced by the {\it Chandra} dither (see
the text for a detailed discussion).
\end{figure}
\clearpage

\begin{table*}
\begin{center}{Table 1: {\it Chandra} Observations of V407 Vul}
\begin{tabular}{cccccc} \\
\tableline
\tableline
 &  OBSID     &  Instrument    &  Start UTC    &  Stop UTC  &  Exp (ksec) \\
\tableline
1 & 300095  &  ACIS-S (CC-mode)  & Feb 18, 2003:11:16:30 & Feb 18, 
2003:15:23:39 & 13.4 \\
2 & 300095  &  ACIS-S (CC-mode)  & Feb 19, 2003:07:49:17 & Feb 19, 
2003:17:58:49 & 35.4 \\
3 & 300096  &  ACIS-S (CC-mode)  & Nov 11, 2003:19:33:05 & Nov 11, 
2003:23:06:22 & 11.2 \\
4 & 300111 &   ACIS-S (CC-mode)  & Dec 27, 2003:02:36:24 & Dec 27, 
2003:07:30:14 & 15.4 \\
5 & 300112 &   ACIS-S (CC-mode)  & Jan 31, 2004:21:20:12 & Feb 01, 
2004:02:05:30 & 15.1 \\
\tableline
\end{tabular}
\end{center}
\end{table*}
\clearpage

\begin{table}
\begin{center}
{Table 2: Timing Solution for V407 Vul}
\end{center}
\begin{center}
\begin{tabular}{cc} \\
\tableline
\tableline
 Model Parameter &  Value \\
\tableline
 $t_0$ (TDB) & MJD 49257.533373137 \\
$\nu_0$ (Hz) & 0.0017562462(2)  \\
$\dot\nu$ (Hz s$^{-1}$) & $7.0(8) \times 10^{-18}$ \\ 
\tableline
\end{tabular}
\end{center}
\end{table}
\clearpage


\end{document}